\documentclass[rnote]{aa} % for the research notes
\usepackage{graphicx}
\usepackage{txfonts}

\def\kms{~km\,s$^{-1}$}

\begin{document}

 \title{Discovery of a [WO] central star
 in the planetary nebula \mbox{Th 2-A}
 \thanks{Based on data collected at (i) the Complejo Astron\'omico
El Leoncito (CASLEO), which is operated under agreement between the
Consejo Nacional de Investigaciones Cient\'{i}ficas y T\'ecnicas de
la Rep\'ublica Argentina y Universidades Nacionales de La Plata,
C\'ordoba y San Juan, Argentina; (ii) the 6.5\,m Magellan Telescopes
at Las Campanas Observatory, Chile; (iii) the 8\,m Gemini South
Telescope, Chile.}}

 \author{W. A.  Weidmann \inst{1,2},
        R. Gamen \inst{2,3},
        R. J. D\'{i}az \inst{3,4}\and
    V. S. Niemela \inst{5}}

 \offprints{W. Weidmann}

 \institute{Observatorio Astron\'omico C\'ordoba,
              Universidad Nacional de C\'ordoba, Argentina\\
              \email{walter@mail.oac.uncor.edu}
              \and Consejo Nacional de Investigaciones 
                   Cient\'ificas y T\'ecnicas, Argentina
              \and Complejo Astron\'omico El Leoncito, Avda. Espa\~na 1412 Sur, 
                   San Juan, Argentina. 
              \and Gemini Observatory, AURA, La Serena, Chile
              \and In memoriam (1936-2006) \\  }

% \abstract{}{}{}{}{}
\abstract
{About 2500 planetary nebulae are known in our Galaxy but only 224 have central stars
with reported spectral types in the Strasbourg-ESO Catalogue of Galactic Planetary Nebulae
(Acker et al. 1992; Acker et al. 1996).}
{We have started an observational program aiming to increase the
number of PN central stars with spectral classification.}
{By means of spectroscopy and high resolution imaging,
we identify the position and true nature
of the central star. We carried out low resolution spectroscopic
observations at CASLEO telescope, complemented with medium
resolution spectroscopy performed at Gemini South and Magellan
telescopes.}
{As a first outcome of this survey, we present for the first time
the spectra of the central star of the PN  \object{Th 2-A}.
These spectra show emission lines of ionized C and O, typical in
Wolf-Rayet stars.} 
{We identify the position of that central star, 
which is not the brightest one of the visual central pair.
We classify it as of type [WO 3]pec, which is
consistent with the high excitation and dynamical age of the nebula.} {} {}


\titlerunning{Discovery of a [WO] central star in the PN Th2-A}
\authorrunning{Weidmann et al.}

 \maketitle

\section{Introduction}

%\textbf{

The central stars of planetary nebulae (CSPN)
play an important role in the understanding
of post-AGB stellar evolution.
Planetary Nebulae (PNs)
are the relics of intermediate-mass stars, and
they are important probes of stellar evolution, stellar Populations, and cosmic
recycling.
Understanding the evolutionary status of particular classes
of CSs is relevant in a broader astrophysical sense.

%Indeed, it is expected that different PNs and CSs would
%have different masses, ages, chemistry,
%progenitors, or companions (or a combination of these factors),
%and would contribute in different ways to the
%galaxy or population to whom they belong.
%Their relevance,
%then, propagates into the wider field of galactic
%evolution and extragalactic population, which in turn has
%an impact on some astrophysical applications
%where PNs are important components.

About 60\% of the PNe listed in the
Strasbourg - ESO Catalogue of Galactic Planetary Nebulae
(Acker et al. 1992; Acker et al. 1996)
include data concerning the CSPN,
but only 30\% of these actually have spectral classification.
Among the 224 spectral types reported in that catalog, there are 75 stars
presenting emission lines typical of Wolf-Rayet (WR) stars
(Acker \&  Neiner 2003, hereafter AN03).

%Those CSPNe whose optical spectra are dominated by
%strong emission lines are termed as [WR] stars,
%in analogy to the massive Population I ones.
%Early papers (Perek \& Kohoutek, 1967)
%have split the WR-CSPN into two subclasses,
%the Nitrogen [WN] and the Carbon [WC] sequence,
%according to the element dominance.
%Only one PN with a [WN] star
%in its nucleus (Morgan et at. 2003) has been detected so far.
The [WO] type (Oxygen sequence) was
introduced by Smith \& Aller (1969)
to classify a group of 7 CSPN showing an optical
WR-like spectrum with strong
\ion{O}{vi} emission lines. Subsequently,
Barlow \& Hummer (1982) defined
WO stars as a new WR subtype. The most prominent
feature of these objects is the strong emission
of the \ion{O}{vi}
$\lambda\lambda$3811-34 \AA\, doublet, 
%which is also
%present, though with lower intensity, in many
%WC4 stars and one WN star (\object{HD 104994}) and is barely
%detected in the spectra of other WR stars. 
other strong emission lines in the spectra of
[WO] stars are the \ion{C}{iv} + \ion{He}{ii} $\lambda\lambda4658-86$ \AA \
blend and \ion{C}{iv} $\lambda\lambda5801-12$ \AA \ doublet, features which
are also exhibited strongly by early WC stars. 
%On the other hand, all the known WO stars show
%extremely high values of stellar-wind velocity ranging between
%4200 \kms and  and 5500 \kms
%(Kingsburgh \& Barlow 1991; Polcaro et al. 1991).
At the moment there are only four WO stars known in our Galaxy
(van der Hucht, 2001; Drew et al. 2004) and about twenty [WO]
(AN03; Gesicki et al. 2006).
%Although there is no clear difference between the optical
%spectra of [WO] and WO stars, quantitatively
%WOs tend to be cooler and to show broader spectral
%emission lines than [WO] stars (Polcaro et al. 1997).

%Although many PNe with very high
%excitation classes (EC) are known,
%and their spectra are available, most of their
%central stars have been disregarded so far in spectral
%analyses because they are very faint in the
%optical wavelength range.
%Thus, numerous CSPNe remain without a spectral classification.
%It is of paramount interest to investigate this kind of stars and their
%surrounding nebulosities, in order to improve
%our knowledge of the final stage of the stellar evolution.

We have started an observational program at CASLEO
with the goal of studying the
spectra of hitherto unknown CSPNe. In this paper we present one of our first results,
i.e. the discovery of a new [WO] star.
%In the next section we describe the observational technique,
%in Section 3, we describe some physical and
%kinematical characteristics of the nebula around the WOs,
%and present our results related with the spectrum of the CSPN.

\section{Observations}
 \label{obser}

The initial observations 
%that originated this discovery 
were carried out on March 21, 2006, using the
REOSC spectrograph attached to the 2.15\,m telescope at CASLEO,
Argentina. These spectra were taken with a
300~line mm$^{-1}$ grating, which provides a dispersion of
3.4~\AA\,px$^{-1}$, and a wavelength range of
$3500$--$7000$~\AA\, on a Tek 1024$\times$1024 pixels CCD.
%(Weidmann, W. \& Niemela, V. 2006). 

\begin{figure}
 \centering
 \includegraphics[scale=0.42]{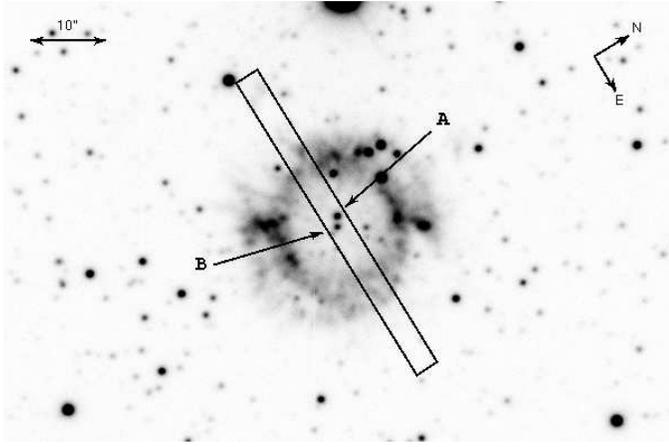}
 \caption[]{The PN Th 2-A as seen with GEMINI-South, filter Ha\_G0336, 
            with a seeing of 0.7 arcsec.
            The orientation and size of the CASLEO slit is indicated.
            The star labeled B is
            the true and newly identified CS.}% of Th 2-A.}
 \label{foto}
\end{figure}

As will be noted in Section~\ref{resultados}, the CS of the PN Th 2-A
has a visual companion separated by 1\farcs4 (See Fig.~\ref{foto}),
which certainly is marginally resolved from CASLEO.
Then, the slit was centered on the stars labeled A and B in
the Fig.~\ref{foto},
in the E--W direction with a slit width of 3\arcsec.
To be sure 
of the sources identifications
%that the B star was marginally present
in the (A+B) CASLEO
spectrum, we observed the A and B stars with the Gemini Multi-Object
Spectrograph (GMOS) at Gemini South, Chile (Proposal
GS-2007B-Q-251). A long-slit of 0.5 arcsec was placed with the
orientation of the line joining the A and B stars. A grating of 
600~line mm$^{-1}$
was used which provided a spectral
coverage of 3630--6440 \AA\ and a dispersion of $\sim 0.45$ \AA\,px$^{-1}$.

N. Morrell provided us an additional spectrum obtained with the Baade Magellan
6.5-m telescope, at Las Campanas Observatory (LCO) in Chile,
%during a 0\farcs7-of-seeing night. 
during a night with 0.7 arcsec seeing.
The Inamori Magellan Areal Camera and Spectrograph
(IMACS) was used in its short camera mode, and the grism 200-15.0 as dispersor.
The observation resulted in a splitted spectrum with a 
wavelength coverage of
3800-6610 \AA\ and 6700-9430 \AA\ and a dispersion of 1.2 and 1.3 \AA\ px$^{-1}$
for the blue and red part respectively.

\section{Results and Discussion}
\label{resultados}

The central star of Th\,2-A (object A in Fig.~\ref{foto})
was described by Kerber et al. (2003)
as being the ``well defined photometric center'' of the nebula (PHOT in their
identification class).
%They used the second epoch Digital Sky Survey (DSS-II) images which have
%a spatial resolution of 1\arcsec~pix$^{-1}$).
%object A in Fig.~\ref{foto}.
Ciardullo et al.\ (1999) determined its Johnson $V$-band magnitude
$V$ = $17.08$ 
%and its $V-I$ color (0.71) 
using
HST images. They included this PN in their survey for resolved companions,
because they noted a nearby companion (star B), which resulted to have a
$\sim 50\%$ probability of being a chance superposition.
%0.38 and 1.63 magnitudes fainter than the A star
%in the HST photometric filters F555 and F814 respectively, thus resulting 
On the other hand,
we did not find any spectroscopic study reported in the bibliography,
related with the CSPN of Th\,2-A.

\begin{table}
 \caption[]{Emission lines identified in the LCO spectrum of the central star of
            the PN Th 2-A over the 3520-9430\AA\ wavelength range.
            First column shows the central wavelength measured
            in our spectrum, when a deblend was not possible
            the centroid of the blend is indicated.
            The marks ":" and "n.m." refers to uncertain values
            and values no measured respectively.}
 \label{star}
 \centering
 \setlength{\tabcolsep}{0.5mm}
 \begin{tabular}{l c c c }
 \hline\hline\noalign{\smallskip}
 Line [\AA]   & Ion & FWHM [\AA] & $W_\lambda $ [\AA] \\
 %\AA   &     &  \AA  & \AA \\
 \noalign{\smallskip}\hline\noalign{\smallskip}
 3822  & \ion{O}{vi}               &   80 &  580  \\
 4118  & \ion{O}{v}                &   31 &  67   \\
 4222  & \ion{C}{iv} ?             &  58  & 22   \\
 4342  & \ion{He}{ii}, \ion{C}{iv} &  40  & 20 \\
 4515  & \ion{O}{vi},\ion{O}{v}    &  101 & 44   \\
 4674  & \ion{C}{iv},\ion{He}{ii}  &  66  &  385    \\
 4856  & \ion{He}{ii}, \ion{C}{iv} &  50  & 9 \\
 5101  & \ion{O}{v} ?              &  9   & 71 \\
 5287  & \ion{O}{vi}               &  54  & 33   \\
 5413  & \ion{C}{iv}, \ion{He}{ii} &  88  & 59 \\
 5468  & \ion{C}{iv}               &  66  & 19  \\
 5590  & \ion{O}{v}                &  92  &  82   \\
 5803  & \ion{C}{iv}               &  86  &  190  \\
 6067  & \ion{O}{viii}, \ion{O}{vii} & 70 & 14  \\
 6197  & \ion{O}{vi}               & 71   & 12 \\
 6383: & \ion{O}{iii} ?            & n.m.     & n.m. \\
 6436: & \ion{He}{ii}, \ion{O}{v}  & n.m.     & n.m.  \\
 6555  & \ion{He}{ii}, \ion{C}{iv} &   79 & 63 \\
 7060  & \ion{C}{iv}               &  88  & 44 \\
 7567: & ?                         & n.m. & n.m.   \\
 7725  & \ion{C}{iv}, \ion{O}{vi}  & 95   & 158 \\
 8844: & ?                         & n.m. & n.m. \\
 9111: & \ion{O}{iii} ?            & n.m. & n.m. \\
 \hline
 \end{tabular}
\end{table}

\begin{figure*}
 \centering
 \includegraphics[width=14.0cm]{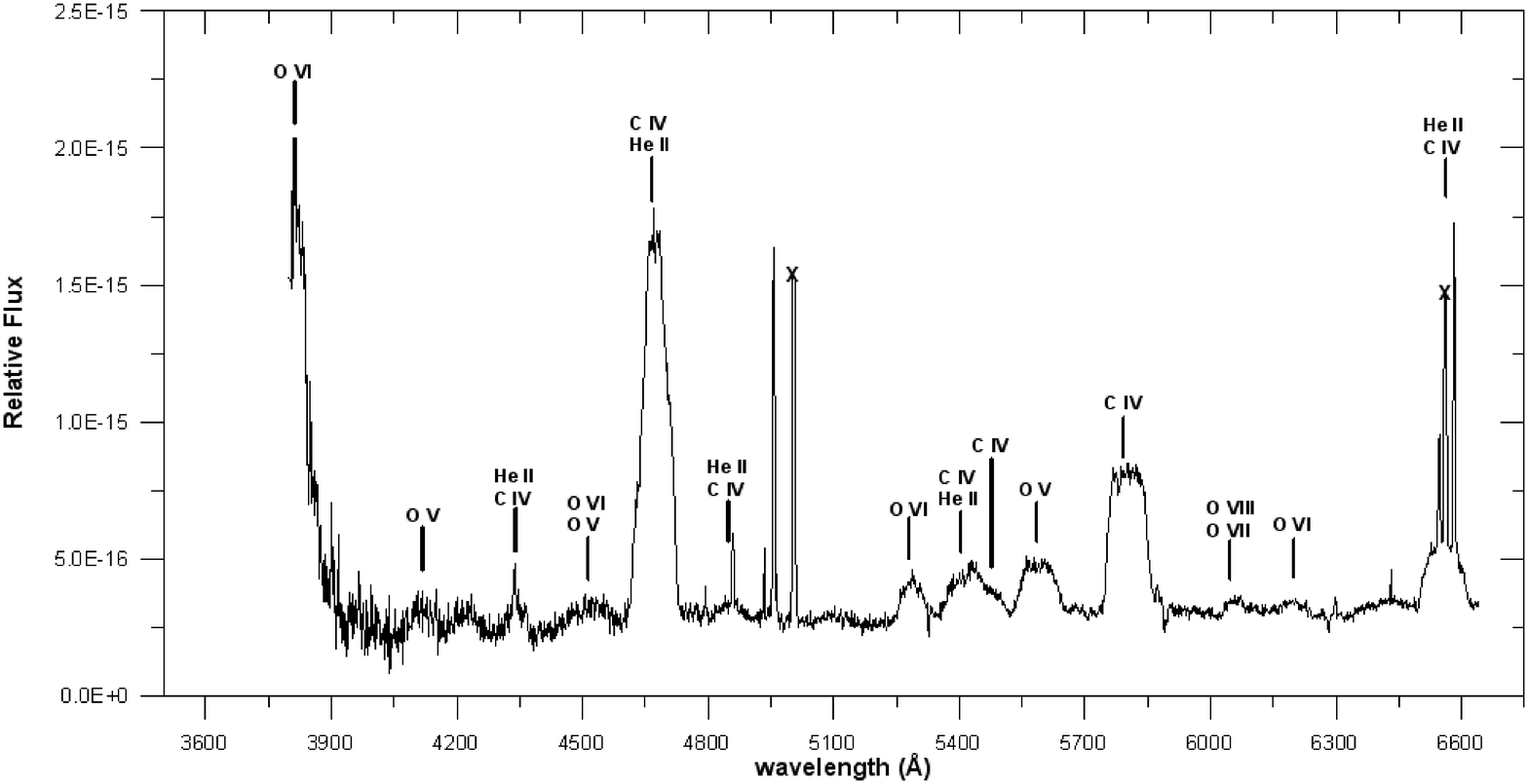}
 \includegraphics[width=13.5cm]{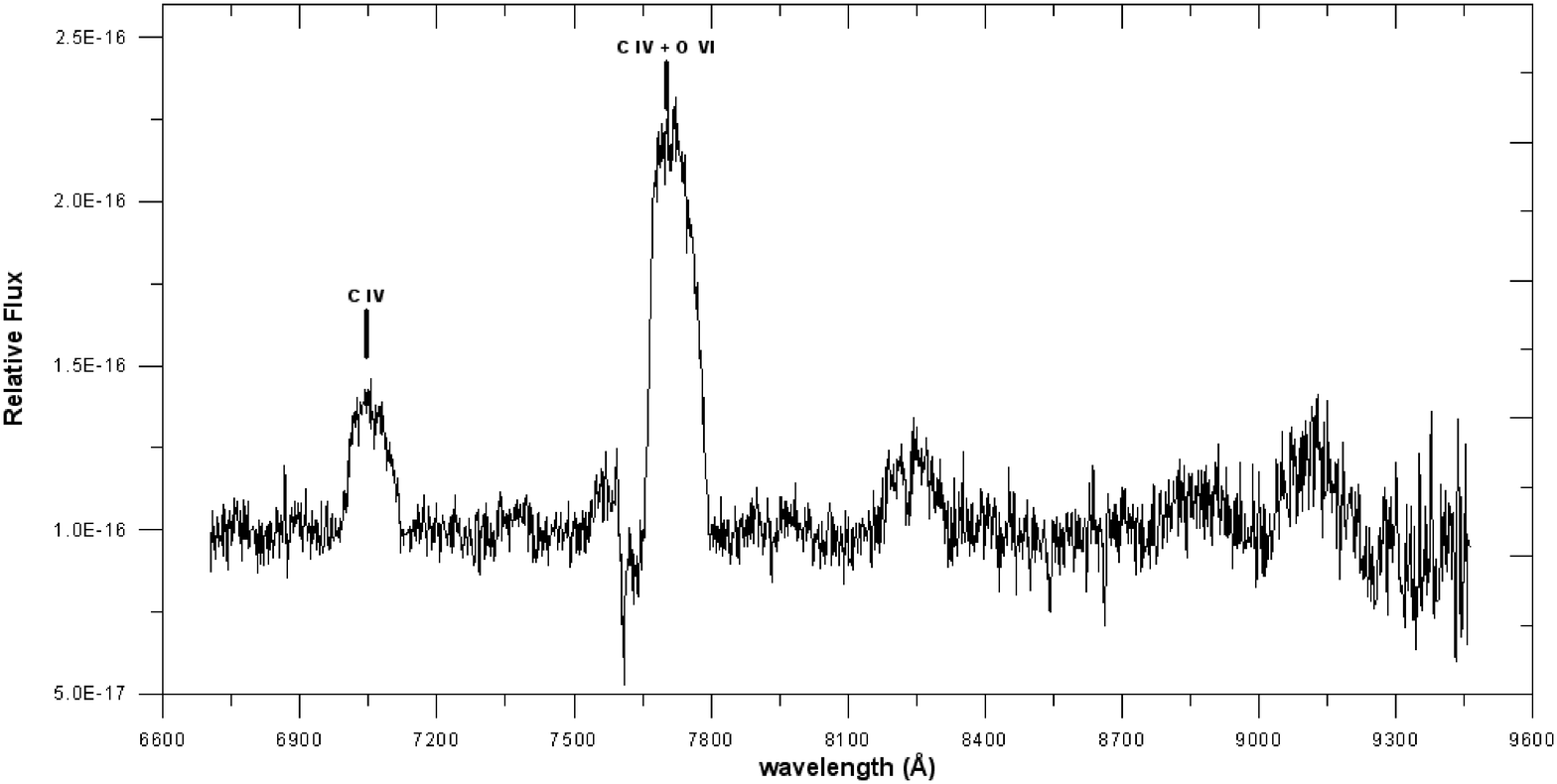}
 \caption[]{Blue and red
            spectra of Th 2-A obtained in December 2007 using LCO.
            The most intense nebular lines
            have been truncated (``X''-labeled)
            in order to show more details of the spectra.
            Note the broad lines from the central
            star at 3822 \ion{O}{vi}, 4670 \ion{C}{iv} and 5805 \ion{C}{iv}.}
 \label{esp}
\end{figure*}

The inspection of our (A+B) CASLEO spectrum revealed that the CSPN
Th2-A was one of the rare [WO] stars, thus claiming our attention.
Searching for published information about this star, we found 
the above cited Ciardullo et al.'s  work 
%and retrieved the higher spacial-resolution
and retrieved higher spatial resolution
HST-WFPC2 images from the MAST archive.
We noted that certainly the B star 
%(also labeled in Fig.~\ref{foto})
seems more centered in the nebula than A. Moreover, we
performed a preliminary aperture-photometry, %on the HST images,
and estimated that the B star is bluer than A.
%(also labeled in Fig.~\ref{foto}).
But, in order to get confidence about the true identification of the
central star of the PN, we obtained a high spatial resolution
spectrum along the A and B stars (see fig~\ref{foto}). Then, the
GMOS-S spectrum (see observation description) confirmed that the B
star is in fact the CSPN and A star is a late type one.
These high spatial resolution images and spectra will 
be presented and discussed elsewhere in a thorough analysis
of the whole planetary nebula system.

In our optical spectra of star B, we identified broad and intense
\ion{C}{iv} and \ion{O}{vi} emission lines, that are typical of WR
stars, thus confirming that this spectrum 
was originated in the nucleus of the PN. 
We identified other faint emission lines of \ion{He}{ii}, \ion{O}{v},
and \ion{O}{vii} 
(see Table~\ref{star} for a list of all 
the identified features),
but some of the faintest ones remained
unidentified.
% as no suitable lines were found in the NIST Atomic
%Spectra Database Lines Data (Ralchenko et al. 2007). 
We specially
noted three emission lines which we identify as \ion{O}{v}
$\lambda5590$, \ion{O}{vi} $\lambda5290$, and \ion{C}{iv}
$\lambda5440$, which are usually observed in WO-type objects. 
%The spectrum is similar to those of other Galactic [WO] CSPNe (Acker \&
%Neiner, 2003), e.g. \object{PNG 324.0+03.5} and \object{PNG 011.9+04.2}. 
In Fig. \ref{esp}, we show the LCO spectrum of Th 2-A
%compared with the one of the [WO3] star IC\,1297 (Acker \&  Neiner 2003), and
and identify the emission lines detected in it.
%Spectra of IC\,1297 was obtained using the same instrumental
%configuration than the one for Th 2-A (see Section~\ref{obser}).
We show this spectrum because it covers the largest spectral range.

We performed a quantitative classification of the spectrum
of the CSPN of Th 2-A
using the criteria described by AN03,
based on dereddened line intensities
ratios and FWHM of emission lines, we obtained that most of the emission lines
strongly indicate a [WO3]-type classification, but the FWHM of the \ion{C}{iv}
$\lambda$ 5806 is similar to the one measured in [WO4]pec-type stars.
Then, we suggest a [WO3]pec spectral classification for the CSPN of Th 2-A.
A note of caution has to  be added, 
the line ratios indicate a type between [WO2]
and [WO4] and the intrinsic scatter in the classification scheme is 
also non-negligible as pointed out by AN03.

%Whit our data from CASLEO we computed some nebular parameters:
%$N_e(S^+)=900$ cm$^{-3}$, 
%$T_e(O^{++})=9900K$ and
%\textit{ EC}= 6.5 
%(computed using the method described
%by Dopita \& Meatheringham, 1990),
%wich are consistent with that 
%determinated by
%Kingsburgh \&  Barlow, 1994.
%From the high excitation of the nebula,
%In addition,
%On the other hand,
%Preite-Martinez (\cite{pm}) estimated, for this
%CSPN, a very high superficial temperature of
%\textit{$T_*$}$=1.57\times 10^5$~k.
%All this parameters are consistent with that
%obtained by Acker \& Neiner (2003),
%from a sample of  [WO].

Th 2-A is an old PN with a dynamical age of 7~kyr estimated 
in a very simplified way, i.e. ignoring 
the effects of the velocity gradient and the acceleration over time, and
assuming a distance of $2.07$ Kpc (Phillips 2004), an angular
size of $27\farcs3$ (Tylenda et al. \cite{ty}) and an expansion velocity
\textit{V}$_{\mathrm{exp}}=18 $\kms\ (Meatheringham et al. \cite{me}).
%According to the
%evolutionary sequence for H-deficient CSPN (Acker \&  Neiner, 2003)
%an early WR-type could be expected as its central star.
%MIENTRAS LEO EL PAPER MENCIONADO INTENTO ESTO:
%Although, the estimations of
%kinematical ages in PN are still matter of debate (Schonberner et al. 2005),
The estimations of kinematical ages in PN are still 
matter of debate (Schonberner et al. 2005), 
but the classification of the CSPN fits well in the evolution sequence 
i.e. post-AGB $\rightarrow$ [WC11] to [WC4] $\rightarrow$ [WO4] to [WO1].
Moreover, the hot spectral-type estimated for the CSPN Th 2-A
is consistent with the high excitation state of the nebula
(Kingsburgh \& Barlow, 1994).

%And also explaines the morphology found by
%Phillips \& Ramos-Larios (2008) from 
%Mid-Infrared {\sc spitzer} images, 
%i.e. the central hole becomes more prominent 
%as wavelength increases.

%\section{Conclusions}

%We present for the first time the optical spectrum of the CSPN
%\mbox{Th 2-A} and confirm that its central star is the one
%previously known as star B (labeled in Fig.~\ref{foto}).

%Considering the AN03
%criteria for classification of CSPNe, we propose
%a [WO3]pec spectral type, which agrees with
%the facts that
%\begin{itemize}
%  \item According to Girard et al. (2007),
%our derived parameters from the plasma diagnostic
%(low \textit{Ne}, high \textit{Te} and \textit{EC})
%are consistent with an early [WO] star.
%dynamical age of the nebula indicates that
%it is an old object, in agreement with the
%evolution sequence of CSPNe, i.e. post-AGB $\rightarrow$ [WC11] to [WC4]
%$\rightarrow$ [WO4] to [WO1].
%  \item The early [WO] spectral type is compatible with the
%very high $T_*$ determinated by
%Preite-Martinez et al. (1989).
%\end{itemize}

Finally, we want to remark that it is important to perform
%detailed observations of this star, i.e.
%best S/N and resolution,  to
%improve our knowledge about this
%kind of objects, which indeed belong
%to a very small group of stars.
multi-epoch observations that will help to
determine its binary nature (if any).
Certainly, there is a surprising lack of
surveys that can detect binaries, letting
the actual PN binary fraction as unknown,
thus some theories that propose that binary
interactions have an important role can not
be tested (e.g. \cite{moe06}).
Moreover, a full quantitative analysis using stellar 
atmospheric modeling will be worth while.

\begin{acknowledgements}

The CCD and data acquisition system at CASLEO has been 
financed
by R. M. Rich trough U. S. NSF grant AST-90-15827. This work
has been partially supported by Concejo de Investigaciones
Cientif\'{i}cas y T\'ecnicas de la Rep\'ublica Argentina (CONICET).
We acknowledge Guillermo Bosch for comments that helped to improve
the paper, and thank to Nidia Morrell for kindly obtaining a
spectrum for us. Some of the data presented in this paper were
obtained from the Multimission Archive at the Space Telescope
Science Institute (MAST). STScI is operated by the Association of
Universities for Research in Astronomy, Inc., under NASA contract
NAS5-26555. Support for MAST for non-HST data is provided by the
NASA Office of Space Science via grant NAG5-7584 and by other grants
and contracts. 
%The Gemini Observatory is operated by the Association
%of Universities for Research in Astronomy, Inc., under a cooperative
%agreement with the NSF on behalf of the Gemini partnership: NSF
%(USA), PPARC (United Kingdom), NRC (Canada), ARC (Australia),
%CONICET (Argentina), CNPq (Brazil) and CONICYT (Chile).
%
Based on observations obtained at the Gemini Observatory, which 
is operated by the Association of Universities for 
Research in Astronomy, Inc., under a cooperative agreement 
with the NSF on behalf of the Gemini partnership: 
the National Science Foundation (United States), the Science 
and Technology Facilities Council (United Kingdom), 
the National Research Council (Canada), CONICYT (Chile), 
the Australian Research Council (Australia), 
Minist\'erio da Ci\^encia e Tecnologia (Brazil) and Ministerio de Ciencia, 
Tecnolog\'{i}a e Innovaci\'on Productiva (Argentina).

\end{acknowledgements}


\begin{thebibliography}{}


%  \bibitem[1989]{ack89} Acker, A., Jasniewicz, G., Koeppen, J., Stenholm, B.
%      1989, A\&AS, 80, 201

  \bibitem[1996]{ack2} Acker, A., Ochsenbein, F., Stenholm, B., et al.
      1992, 1996, Strasbourg-ESO catalogue of galactic planetary nebulae, ESO

  \bibitem[2003]{ack} Acker, A. \&  Neiner, C.
      2003, A\&A, 403, 659 (AN03) \label{ack}

  \bibitem[1982]{bh} Barlow, M. J. \& Hummer, D. C.
      1982, Wolf-Rayet Star: Observation, Physics, Evolution.
      In: de Loore C. W. H. \& Willis A. J. (eds.) Proc.
      IAU Symp. 99, Reidel, Dordrecht, p. 387

%  \bibitem[1999]{benj} Benjamin, R. A., Skillman, E. D., Smits, D. P.
%      1999, ApJ, 514, 307

%  \bibitem[1971]{b} Brocklehurst, M. 1971, MNRAS, 153, 471

  \bibitem[1999]{hst} Ciardullo, R., Bond, H. E., Sipior, M. S.,
     Fullton, L. K., Zhang, C. Y., Schaefer, K. G.
      1999, AJ, 118, 488 \label{hst}

%  \bibitem[1987]{dero} De Robertis, M. M., Dufour, R. J., Hunt, R. W.
%      1987, JRASC, 81, 195

%  \bibitem[1990]{dome}  Dopita, M. A. \& Meatheringham, S. J.
%      1990, ApJ, 357, 140

  \bibitem[2004]{dbupwpmp} Drew, J. E., Barlow, M. J., Unruh, Y. C., Parker, Q. A.,
       Wesson, R., Pierce, M. J., Masheder, M. R. W., Phillipps, S.
      2004, MNRAS, 351, 206

%  \bibitem[2006]{gzaggw} Gesicki, K., Zijlstra, A. A., Acker, A.,
%      G\'orny, S. K., Gozdziewski, K., Walsh, J. R.
%      2006, A\&A, 451, 925

  \bibitem[2007]{gka} Girard, P., K$\ddot{o}$ppen, J., Acker, A.
      2007, A\&A, 463, 265

  \bibitem[2001]{v} van der Hucht, K. A.
      2001, NewAR, 45, 135  \label{v}

  \bibitem[2003]{kgw} Kerber, F., Mignani, R.P., Guglielmetti, F., Wicenec, A.
      2003, A\&A 408, 1029

%  \bibitem[1995]{kff} Kingdon, J. \& Ferland, G. J.
%      1995, ApJ, 442, 714

%  \bibitem[1995]{kb91} Kingsburgh, R. L.\&  Barlow, M. J.
%      1991, IAU, 143, 101

 \bibitem[1995]{kb94} Kingsburgh, R. L.\&  Barlow, M. J.
      1994, MNRAS, 271, 257 

%  \bibitem[1995]{kbs} Kingsburgh, R. L.,  Barlow, M. J., Storey, P. J.
%      1995, A\&A,  295, 75

%  \bibitem[1993]{ku} Kunth, D. \& Sargent, W. L. W.
%      1993, ApJ,  273, 81

  \bibitem[1988]{me} Meatheringham, S. J., Wood, P. R., Faulkner, D. J.
      1988, ApJ, 334, 862 \label{me}

  \bibitem[Moe \& De Marco, 2006]{moe06} Moe, M. \& De Marco, O.
      2006, ApJ, 650, 916

%  \bibitem[2003]{morg} Morgan, D. H., Parker, Q. A., Cohen, Martin
%      2003, MNRAS, 346, 719 \label{morg}

%  \bibitem[2001]{olsk} Olive, K. A. \& Skillman, E. D.
%      2001, NewA, 6, 119 \label{olks}

%  \bibitem[1967]{pk} Perek, L. \& Kohoutek, L.
%      1967, Catalogue of Galactic Planetary Nebulae, Prague, Czech. Acad. Sci.

%  \bibitem[2008]{prl} Phillips, J. P. \& Ramos-Larios, G.,
%      2008, MNRAS, 383, 1029

  \bibitem[2004]{p}  Phillips, J. P.
      2004, MNRAS, 353, 589

%  \bibitem[1991]{pgmnr} Polcaro, V. F., Giovannelli, F., Manchanda, R. K., Norci, L., Rossi, C.
%      1991, A\&A, 252, 590

%  \bibitem[1997]{prvn} Polcaro, V. F., Rossi, C., Viotti, R., Norci, L.
%      1997, A\&A, 318, 571

%  \bibitem[1989]{pm} Preite-Martinez, A., Acker, A., Koeppen, J., Stenholm, B.
%      1989, A\&AS, 81,309  \label{pm}

%  \bibitem[2007]{nist}
%Ralchenko, Yu., Jou, F.-C., Kelleher, D.E., Kramida, A.E., Musgrove, A., Reader, J., Wiese, W.L., and Olsen, K. (2007). NIST Atomic Spectra Database (version 3.1.3), [Online]. Available: http://physics.nist.gov/asd3 [2008, January 22]. National Institute of Standards and Technology, Gaithersburg, MD.

%  \bibitem[1979]{s}  Seaton, M. J.
%      1979, MNRAS, 187, 785

%  \bibitem[1995]{shd} Shaw, R. A. \& Dufour, R. J.
%      1995, PASP, 107, 896


 \bibitem[2005]{sch}Schonberner, D., Jacob, R., Steffen, M.
         2005, A\&A, 441, 573




  \bibitem[1969]{sa} Smith, L. F. \& Aller, L. H.
      1969, ApJ, 157, 1245

%  \bibitem[1983]{sb} Stone, R. P. S. \& Baldwin, J. A.
%      1983, MNRAS, 204, 347  \label{sb}

%  \bibitem[1962]{the} The, P. S.
%      1962, CoBos 17   \label{the}

  \bibitem[2003]{ty} Tylenda, R., Si\'odmiak, N., G\'orny, S. K., Corradi, R. L. M., Schwarz, H. E.
      2003, A\&A, 405, 627 \label{ty}

%  \bibitem[2006]{wV} Weidmann, W. \&  Niemela, V.
%      2006, BAAA, 49, 198    \label{wV}

%  \bibitem[1967]{wh} Westerlund, B. E. \& Henize, K. G.
%      1967, ApJS, 14, 154    \label{wb}


\end{thebibliography}
\end{document}